%Paper: hep-th/9208016
%From: IMBIMBO@genova.infn.it
%Date: Wed, 5 Aug 1992 15:13:07 +0200 (WET-DST)

%
%         TO BE PRINTED USING THE MACRO HARVMAC
%

\input harvmac.tex

\def\np{Nucl. Phys.}
\def\pl{Phys. Lett.}

\def\cmp{ Comm. Math. Phys.}
\def\ijmp{Int. J. Mod. Phys.}
\def\mpl{Mod. Phys. Lett.}
\def\ap{Ann. Phys.}

\def\Ph{{\cal M}}
\def\Pho{{\cal M}^{'}}
\def\Phai{{\cal M}_i}
\def\Phc{{\cal M}_1}
\def\Phn{{\cal M}_2}
\def\Phl{{\cal M}_3}
\def\Ns{{\cal N}}
\def\Phcs{\Phc/\Ns}
\def\Phns{\Phn/\Ns}
\def\Phs{\Ph/\Ns}

\def\Pol{{\cal T}}
\def\G{$SL(2,R)$}
\def\Go{$SO(1,2)$}
\def\Lo{{\cal L}^{(k)}_{o}}

\def\wf{\Psi(z)}
\def\wfa{\psi^{\alpha}(z)}

\def\wfaN{\psi^{\alpha}_N (\tau;z)}

\def\wfN{\Psi_N (\tau;z)}
\def\Hc{ {\bf \cal H}_{\Phcs}}
\def\Hn{ {\bf \cal H}_{\Phns}}
\def\Hcpm{ {\bf \cal H}_{\Phcs}^{\pm}}
\def\Hcm{ {\bf \cal H}_{\Phcs}^{-}}

\def\HCS{ {\bf \cal H}_{CS}}
\def\Rpq{{\bf R}^{(p;q)}}
\def\Rqp{{\bf R}^{(q;p)}}
\def\RCS{{\bf R}_{CS}^{(s/r)}}

\def\Sqp{S^{(q;p)}}

\def\rab{{\cal R}^{(\vartheta_a ,\vartheta_b)}_{p/q}}
\def\roo{{\cal R}^{(0,0)}_{p/q}}

\def\rhalf{{\cal R}^{(1/2,1/2)}_{p/q}}
\overfullrule=0pt
\parskip=4pt plus 2pt minus 2pt
\hsize=15truecm
\vsize=22truecm
{\nopagenumbers
\font\bigrm=cmb10 scaled\magstep1
\rightline{GEF-TH 5/1992}
\ \medskip
\centerline{\bigrm SL(2,R) CHERN-SIMONS THEORIES WITH RATIONAL CHARGES}
\centerline{\bigrm AND 2-DIMENSIONAL CONFORMAL FIELD THEORIES}
\vskip 1truecm
\centerline{CAMILLO IMBIMBO}
\vskip 5pt
\centerline{\it I.N.F.N., Sezione di Genova}
\centerline{\it Via Dodecaneso 33, I-16146 Genova, Italy}
\vskip 8pt
\vskip 2truecm
\centerline{ABSTRACT}
\noindent We present a hamiltonian quantization of the $SL(2,R)$
3-dimensional Chern-Simons theory  with fractional
coupling constant $k=s/r$ on a space manifold with torus topology
in the ``constrain-first'' framework.
By generalizing the ``Weyl-odd'' projection
to the fractional charge case, we obtain multi-components holomorphic wave
functions whose components are the Kac-Wakimoto characters of the
modular invariant admissible representations of ${\hat A}_1$ current
algebra with fractional level.
The modular representations carried by the quantum Hilbert space satisfy
both Verlinde's and Vafa's constraints coming from conformal field theory.
They are the ``square-roots'' of the representations associated to the
conformal $(r,s)$ minimal models.
Our results imply that Chern-Simons theory with $SO(2,2)$
as gauge group, which describes $2+1$-dimensional gravity with negative
cosmological constant, has the modular properties of the
Virasoro discrete series. On the way, we show that the 2-dimensional
counterparts of Chern-Simons $SU(2)$ theories with half-integer charge $k=p/2$
are the modular invariant $D_{p+1}$ series of ${\hat A}_1$ current
algebra of level $2p-2$.
\ \vfill

\leftline{GEF-TH 5/1992}
\leftline{March 1992}
\eject     }

%\nopagenumbers
\pageno=1
\beginsection 1. Introduction

Three-dimensional Chern-Simons topological gauge theory
\ref\witjones{E. Witten,\cmp\ {\bf 121} (1988) 351.}\
with \G\ as gauge group has attracted considerable interest for various reasons
\ref\hverlinde{H. Verlinde, Princeton Preprint, PUTP-89/1140.}%
\nref\ehverlinde
{E. Verlinde and H. Verlinde, Princeton Preprint, PUTP-89/1149.}%
\nref\witcentral{E. Witten, IAS Preprint, IASSNS-HEP 89/38.}%
-\ref\natan{D. Bar-Natan and E. Witten, IAS Preprint,  IASSNS-HEP-91/4.}
including its relationship to both 2-dimensional
\ref\polyakov{V.G. Knizhnik, A.M. Polyakov and A.B. Zamolodchikov, \mpl\
{\bf A3} (1988) 819.} and 3-dimensional quantum gravity
\ref\witgravity{E. Witten, \np\ {\bf B311} (1988/9) 46; \np\ {\bf B323} (1989)
113.}.

Chern-Simons gauge theories with non-compact gauge groups
are not expected to present any special pathology as 3-dimensional
quantum field theories. Their hamiltonian being identically zero and
their action being linear in time derivatives,
one expects on general grounds that they
define perfectly unitary quantum theories \ref\witcomplex{E. Witten,
IAS Preprint IASSNS-HEP-89/65, to appear in \cmp }.
Therefore it appears that their 2-dimensional counterparts
cannot be the non-unitary Wess-Zumino-Witten
models on non-compact group manifolds: if 2-dimensional quantum field
theories associated to Chern-Simons theories with non-compact
gauge group do exist, they are likely to represent some, possibly
yet unknown, generalization
of current algebra constructions. Understanding such generalization
is another motivation to study \G\ Chern-Simons gauge theory.

Unfortunately, the extension of the Hamiltonian quantization techniques which
allowed a non-perturbative solution of Chern-Simons theories with compact
gauge groups to theories with non-compact gauge groups is revealed to be
problematic \witcomplex .
Canonical quantization in the holomorphic ``quantize-first'' scheme
\ref\elitzur{S. Elitzur, G. Moore, A. Schwimmer and N. Seiberg,
\np\ {\bf B326} (1989) 108.}%
\nref\bos{ M. Bos and V.P. Nair, \pl\ {\bf B223} (1989) 61; \ijmp\ {\bf A5}
(1990) 959.}%
-\ref\axelrod{
S. Axelrod, S. Della Pietra and E. Witten, J. Diff. Geom. {\bf 33}
(1991) 787.}
has been essential for establishing the correspondence between
3-dimensional Chern-Simons theory with compact gauge groups and
2-dimensional current algebras, but this approach is not viable
for the real non-compact \G\  due to the lack of a gauge invariant
polarization. Analyses based on polarizations which are not gauge invariant
\hverlinde -\ehverlinde\ provided some intriguing information about \G\
Chern-Simons theory, but were difficult to carry out at explicit
and less formal levels and were limited to the case of trivial space topology.
Recent perturbative computations \natan\ have
stressed the substantially novel features that non-compact gauge
groups introduce into the quantization of topological Chern-Simons theories.

In this paper we will present a canonical quantization of Chern-Simons
theory with \G\ as gauge group in the ``constrain-first'' framework
\elitzur ,\ref\imbimbo{ C. Imbimbo, \pl\ {\bf B258} (1991) 353.}. This
approach, being gauge invariant {\it ab initio}, avoids the difficulties
of non-gauge invariant polarizations affecting the ``quantize-first''
method. We will limit ourselves to the case when the ``space''
manifold $\Sigma$ is a 2-dimensional torus; such limitation has
been sufficient to
unravel the underlying 2-dimensional current algebra structure
in the compact case.

The starting point in the ``constrain-first'' approach is the classical
gauge invariant phase space $\Ph$, the space of flat gauge connections on the
space manifold $\Sigma$. Since $\Ph$ is finite-dimensional,
the canonical quantization problem actually has a finite number of
degrees of freedom.
However, the fact that $\Ph$ is not in general a smooth manifold,
makes its quantization rather non-standard.
Even in the
case of compact gauge groups, $\Ph$ has singularities of finite order
which are associated to important quantum-mechanical effects,
such as the ``shift'' of the central charge in the Sugawara
construction for 2-dimensional algebras \imbimbo .
When the gauge group is \G , the singularities of $\Ph$
are of a more general type, as we will shortly see: they play a central role
in the quantization  of the \G\ Chern-Simons theory which
we consider here.

When $\Sigma$ is a torus, the problem of quantizing $\Ph$ is
reduced to the problem of quantizing the moduli space of flat-connections
of an {\it abelian} gauge group \axelrod . This makes the
computation for genus one drastically simpler than
for higher genus, where non-abelian Chern-Simons theory appears to be
vastly more complex than abelian.
On the other hand, the factorization properties of 2-dimensional
conformal field theories suggest that
the torus topology already contains most, if not all, of the
complexities of higher genus. The solution of this apparent paradox
is that $\Ph$ for a torus is {\it almost}  the space of flat
connections of an abelian group, but not quite: it is the space of
abelian flat connections modulo the action of a discrete group whose fixed
points give rise to orbifold singularities. It is only here that
the quantization of non-abelian Chern-Simons theory (with compact
gauge group) for genus one
differs from the computationally trivial abelian case.
Thus, in some sense, the singularities of
$\Ph$ for genus one must encode much of the information about
the theory on higher genus surfaces,
at least for compact non-abelian gauge groups.
That this remains true for non-compact gauge groups like \G\ is
plausible, though yet to be proven.

In comparison with $\Ph_{SU(2)}$, the distinctive feature of the
phase space $\Ph_{SL(2,R)}$ is its non-compactness. Even if one
restricts oneself (as we will essentially do in this paper)
to the sector of $\Ph_{SL(2,R)}$ corresponding to flat connections
which lie in the ``compact'' Cartan subgroup of \G , one has to deal
effectively with the non-compact smooth manifold obtained by deleting the
singularities from the compact non-smooth phase space.
One consequence of the non-compactness of the (effective) phase space
is that the integrality condition on the Chern-Simons charge $k$
\witjones ,\elitzur ,\bos\ disappears. Another consequence
is that infinite-dimensional Hilbert
spaces emerge, in general, upon quantization. However, if one takes $k$
to be {\it rational}, the Hilbert space of quantum states
becomes finite-dimensional
\ref\poly{A.P. Polychronakos, \pl\ {\bf B241} (1990) 37; \ap\ {\bf 203}
(1990) 231.}.
Therefore, it is reasonable to think that
theories with rational $k$ correspond to rational conformal 2-dimensional
field theories, or to some ``deformation'' of them, which, when $k$ becomes an
integer, reduce to
the familiar non-abelian current algebras.
Recently, abelian Chern-Simons theories with rational charge $k=p/q$ have been
investigated because of their (possible) relevance
to the theory of quantum Hall effect and to a new mechanism for
superconductivity \ref\lechner{For a recent review, see:
R. Jengo and K. Lechner, SISSA-ISAS 174/91/EP, to appear on
Physics Reports.}. Wave functions are represented by
$q$-dimensional {\it multiplets} of theta-functions  of level $pq$,
\ref\jengo{R. Jengo and K. Lechner, SISSA Preprint,
SISSA-182-90-EP.},\ref\imb{C. Imbimbo, preprint GEF-TH 9/1991, to appear in the
``Proceedings of the Workshop on String Theory'', Trieste, March 1991,
edited by H. Verlinde.} which can be thought of as describing holomorphic
sections of a holomorphic ``line bundle'' with rational Chern class $p/q$
on the non-compact phase space. When $k$ is an integer, by
projecting the abelian Hilbert space to the ``Weyl'' odd sector,
one obtains the non-abelian wave functions, that is the
Kac-Weyl characters of the integrable current algebra representations
\axelrod ,\imbimbo . In this paper we will show that the
appropriate generalization of such projection to the fractional charge case
leads to the modular invariant Kac-Wakimoto
\ref\vgkac{V.G. Kac and M. Wakimoto, Proc. Nat. Acad. Sci. {\bf 85}
(1988) 4956.} characters of the (non-integrable
and non-unitary) representations of ${\hat A}_1$ current algebra
with fractional admissible level.
We will also discover that the modular representations
acting on the Hilbert space of states of the \G\ Chern-Simons theory
are identical to one of the two factors into which the modular
representations of the conformal minimal models factorize. This
suggests that the 2-dimensional counterpart of the \G\ Chern-Simons
theory might be non-conformal. It also implies that Chern-Simons
theory with gauge group $SL(2,R)\times SL(2,R)\approx SO(2,2)$
and rational charges $(k,{1\over 4k})$ has exactly the modular
properties of the conformal minimal models.
Since $SO(2,2)$ Chern-Simons theory describes $2+1$ dimensional
gravity with negative cosmological constant, it seems therefore supported
Witten's speculation \witgravity\ that gravity in $2+1$ dimensions and
the Virasoro discrete series are related.

\bigskip
\beginsection 2. Geometric Quantization of $\Ph_{SL(2,R)}$

Flat \G\ connections on a torus correspond to pairs $(g_1,g_2)$
 of commuting
\G\ elements, modulo overall conjugation in \G . $\; g_1$ and $g_2$
represent the holonomies of the flat connections around the two non
trivial cycles of the torus. \G\ has a non-trivial $Z_2$ center and
$SO(1,2) \approx SL(2,R)/Z_2 $. Therefore, $\Ph$ is a four-cover of the
space $\Pho$ of \Go\ flat connections, since to each \Go\ flat connection
correspond four \G\ flat connections whose holonomies differ by elements of
the center $Z_2$.
It is convenient to describe $\Ph$ in terms of the simpler $\Pho$.
Let us think of \G\ as the group of unimodular $2 \times 2$ real matrices.
The basic fact of $\Ph$ ( or $\Pho$) is that it is the union of three
``sectors''
\eqn\one{ \Ph = \bigcup_{i=1,2,3} \Phai \; ,}
where the $\Phai$'s, $i=1,2,3$, are the space of \G\ flat connections whose
respective holonomies have two imaginary (and conjugate) eigenvalues ($i=1$),
two eigenvectors with real (and reciprocal) eigenvalues ($i=2$), and
one single eigenvector with unit eigenvalue ($i=3$).

When $i=1$, both holonomies can be simultaneously brought by conjugation
into the compact
$U(1)$ subgroup of \G . Therefore $\Phc \approx T^{(1)}$, the two
dimensional torus. Let us introduce the real coordinates
$(\theta_1,\theta_2)$ for $\Phc$.  In our normalization,
the periodic coordinates $\theta_{1,2}$ lie in the unit real
interval when the gauge group is \Go ; for \G , these take values in the
enlarged interval of length 2.

For $i=2$, the holonomies can be conjugated into a diagonal form.
However, one can still conjugate diagonal holonomies by an element of
the gauge group which permutes the eigenvalues.
Therefore, when the gauge group is \Go ,
$\Phn \approx R^{(2)}/Z_2$. If $(x,y)$ are cartesian coordinates on the
real plane $R^{(2)}$, the $Z_2$ action is the reflection around the origin,
mapping $(x,y)$ onto $(-x,-y)$. If the gauge group is \G , $\Phn$ consists
of four copies of $R^{(2)}/Z_2$.

Finally, when $i=3$, holonomies can be conjugated into an upper triangular
form with units on the diagonal. Conjugation allows one to rescale the
(non-vanishing) elements in the upper right corner by an arbitrary positive
number. Thus, $\Phl \approx S^1$, the real circle.
Being odd-dimensional, $S^1$ cannot be a
genuine non-degenerate symplectic space. In fact, the symplectic form
on the space of flat connections coming from the Chern-Simons action,
when pushed down to $\Phl$ vanishes identically. $\Phl$ represents
a ``null'' direction for the symplectic form of the \G\ Chern-Simons
theory, reflecting the indefiniteness of the \G\ Killing form.
Since $\Phl$ is a {\it disconnected} piece of the
total phase space $\Ph$ (or $\Pho$),
it is consistent to consider the problem of quantizing
$\Phc \cup \Phn$ independently of $\Phl$. After all, modding out by the
``null'' directions (such as those originated by gauge symmetries), is
the common recipe for dealing with degenerate symplectic forms.
Hence, in what follows we concentrate on $\Phc \cup \Phn$, though
it is conceivable that the ``light-like'' sector $\Phl$ merits
further investigation.

To summarize, the space of gauge  flat connections on the torus
(disregarding $\Phl$) looks as follows: a torus ($\Phc$) with planes
($\Phn$) ``attached'' to it at the points $z_s$ in a
discrete set $\Ns \equiv \Phc \cap \Phn $, representing
flat connections with holonomies in the center of the gauge group.
For the \Go\  case, $\Ns$ contains a single point, whose $\Phc$ and $\Phn$
coordinates are $({\theta_1}^{(s)},{\theta_2}^{(s)}) = (0,0)$ and
$(x^{(s)},y^{(s)}) = (0,0),$ respectively. When
the gauge group is \G , $\Ns$ consists of four points,
with  $({\theta_1}^{(s)},{\theta_2}^{(s)}) =(\pm 1,\pm 1)$.
The $\Phn$ planes are ``folded'' by the $Z_2$ reflections around the
points in $\Ns$.
%[The situation is illustrated in fig.1].

The distinctive feature of classical phase space $\Ph$ is that  it ceases to
be a smooth manifold around the points in $\Ns$. The quantization
of the classical phase space $\Ph$ involves considering smooth functions or
smooth sections of appropriate line bundles on $\Ph$, raising the
question of the meaning of ``smooth'' sections on a non-smooth manifold
such as $\Ph$.
Our strategy is to consider first the smooth, non-compact
manifold $\Ph/\Ns$ obtained by deleting the singular points in $\Ns$.
$\Phs$ consists of two disconnected smooth components, $\Phcs$ and $\Phns$.
We will then consider quantizations of $\Phcs$ and $\Phns$ which admit
sections that can be ``glued'' at the points in $\Ns$. The final
Hilbert space will be the span of those ``glued'' sections.
Our ``intuitive'' approach could conceivably be substantiated
with more rigorous methods of algebraic geometry.

We will perform the quantization of $\Phcs$ and $\Phns$
in the holomorphic scheme \elitzur -\axelrod\ since,
as is familiar from the study of the compact Chern-Simons
theory \witjones , $\Ph$ admits
a natural family $\Pol$ of K\"ahler polarizations \ref\woodhouse{
N. Woodhouse, ``Geometric Quantization'', Oxford University Press,
Oxford 1980.}.
$\Pol$ is the Siegel upper complex plane, because the choice of a complex
structure on the 2-dimensional space manifold $\Sigma$
induces a complex structure on the space of connections on $\Sigma$
and, by projection, on $\Ph$.
For $\tau \in \Pol$, let us introduce holomorphic coordinates  on both
$\Phcs$
\eqn\holoc{ z = \theta_1 + \tau \theta_2, \;\;\; {\bar z} = \theta_1 +
{\bar \tau} \theta_2, \;\;\; (\theta_1,\theta_2)\in \Phcs}
\noindent and $\Phns$
\eqn\holon{ z = x + \tau y, \;\;\; {\bar z} = x +
{\bar \tau}y, \;\;\; (x,y)\in \Phns.}

\noindent Then the symplectic form which descends from the Chern-Simons action
with charge $k$
\eqn\azione{S= {k\over 4\pi}\int_{\Sigma \times R^1}
\langle A,dA+{1\over 3}A\wedge A\rangle }

\noindent can be written both on $\Phc$ and $\Phn$ in the coordinates
systems \holoc\ and \holon\ as follows:
\eqn\form{ \omega = {i k\pi \over 2\tau_2} dz \wedge d{\bar z}\equiv
i {\bar \partial}\partial K ,\;\;\; \tau_2 \equiv Im \tau}
\noindent where $K$ is the K\"ahler potential which we choose to be
\eqn\Kahler{ K = {k \pi \over 4\tau_2} \left(z- {\bar z}\right)^2.}

In the context of K\"ahler quantization, the Hilbert space of quantum
states is the span of square integrable holomorphic sections
of a holomorphic line bundle with hermitian structure whose curvature
two-form is the symplectic form $\omega$ in \form .

The quantization of $\Phns$ is rather straightforward. Since $\Phns$ is
not simply connected, the holomorphic wave functions $\psi (z)$
can acquire an
arbitrary phase $e^{2\pi i\vartheta_i}$ when moving around the singular points
$z_i = 0$ of $\Ns$.
The B\"ohm-Aharonov phases $e^{2\pi i\vartheta_i}$ should
be regarded as free parameters of the quantization.
A further two-fold ambiguity of the $\Phns$ quantization stems from the
fact that the gauge invariant $\Phns$ is the quotient of the complex
plane (with the origin deleted) by the action of the reflection
around the origin. Thus, physical wave functions should be invariant
under the action of the unitary operator implementing the reflection around
the origin. Since there are two ways of implementing reflections
according to the ``intrinsic'' parity of the wave
functions, one concludes that the wave functions on each of the four
``sheets'' of $\Phns$ are
\eqn\waven{\psi^{(\vartheta,\pm)}(z) = z^{\vartheta} \chi^{(\pm)}(z),}

\noindent where $\chi^{(\pm)}(z)$ is holomorphic, even (odd) around
the origin, and each choice
of $(\vartheta, \pm)$ is associated to a different quantum Hilbert space
${\cal H}^{(\vartheta, \pm)}_{\Phns}$.

Let us now turn to $\Phcs$.
The crucial difference between quantum mechanics on the non-compact
$\Phcs$ and on the compact torus $\Phc$ originates from the fact that
the homotopy group $\pi_1(\Phcs)$ is {\it non-abelian}:
\eqn\homotopy{ a b = b a \delta, \;\; [a, \delta] =[b, \delta] =0 ,}

\noindent where $a$ and $b$ are the non-trivial cycles of the compact
torus and  $\delta= \prod_i \delta_i$ is the product of the cycles
$\delta_i$ surrounding the
singularities $z_i$ in $\Ns$.
%[Eq. \homotopy for $\Pho$ is represented in Fig.2]
In this case, quantum states are represented by {\it multi-components}
wave functions $\wf = ( \wfa )$, $\alpha = 0,1,...,q-1,$ transforming
in some irreducible unitary, $q$-dimensional representation of the homotopy
group $\pi_1(\Phcs)$.
Let us consider a basis for such  representation which
diagonalizes the $\delta_i$'s. For the representation to be finite-dimensional
and irreducible, the $\delta_i$'s must be represented by {\it rational}
phases.
Moreover, we take all $\delta_i$'s to be the same, since we require
that modular transformations (which mix the singular points in $\Ns$)
act on the Hilbert space of wave functions.
In conclusion we take $ \delta = exp (-2\pi i p/q)$ with $p$ integer,
coprime with $q$.

In the holomorphic quantization scheme, wave functions $\wf$
should be holomorphic and,
in the trivialization corresponding to \Kahler , should have
the periodicity properties of theta functions with fractional
``level'' $k$:
\eqn\transition{ \Psi (z + 2m + 2n \tau ) =
exp ( -2 \pi i k \tau n^2 - 2 \pi i k z n ) a^m b^n \Psi (z), }

\noindent where $a$ and $b$ are $q\times q$ unitary matrices which
provide a representation of homotopy relations \homotopy. Note that
on the compact torus $\Phc$, $a$ and $b$ would be one-dimensional phases
and the consistency (cocycle) condition for the transition functions in
\transition\ would require $2k$ to be an integer \elitzur-\axelrod .
In our case, the consistency condition coming from \homotopy\
relates the Chern-Simons charge $k$ to the monodromy of the wave
functions around the singular points:
\eqn\cocycle{ e^{2\pi i2k}= e^{2\pi i p/q} .}
Therefore, we
restrict ourselves henceforth to the case of $k$ {\it rational}:
\eqn\charge{2k= 2s/r = p/q ,}
\noindent with $s$ and $r$ integers, relatively prime, and $r$ chosen to
be positive. It should be stressed that the
restriction to $k$ rational is motivated by the interest
to investigate the connection between \G\ Chern-Simons
theory and 2-dimensional {\it rational} conformal field theories.
When $k$ is irrational one expects an infinite-dimensional Hilbert space
of holomorphic wave functions: an interesting possibility, which we do not
pursue here.

The holomorphic components $\wfa$ of the wave functions $\wf$ can be
thought of as representing
holomorphic sections of the holomorphic line bundle
$\Lo$ on the non-compact $\Phcs$ with fractional ``Chern-class'' $p/q$.
Locally, a section $\psi$ of $\Lo$ would be given by the $q$-root of a
theta function of level $p$. $\psi$ would have non-trivial mondromy $\delta
=e^{2\pi ip/q}$ around the singular points in $\Ns$, but would be
single-valued when holomorphically extended to a $q$-cover ${\tilde \Phc}$
of the torus $\Phc$. If $\Phc$ is the complex torus with modular parameter
$\tau$, the $q$-cover ${\tilde \Phc}$ is a torus with modular parameter
$q\tau$. The $q$ components $\wfa$ of the wave function $\wf$ should be
identified with the different holomorphic extentions of $\psi$ to ${\tilde
\Phc}$: they should therefore be theta functions of level $q\times p/q=p$
on the torus with modular parameter $q\tau$. We will verify that this
is in fact the case. In the following however we will simply think
of $\wf$ as hoomorphic sections of a vector bundle on the compact
torus $\Phc$ with fibers of dimensions $q$.

It follows from \transition\ that inequivalent holomorphic quantizations
of $\Phcs$ with the same $k$ are in one-to-one correspondence
with classes of inequivalent, unitary and irreducible
representations of the `t Hooft algebra
\eqn\tHooft{ ab = ba e^{2\pi ip/q}.}
\noindent Let us denote by $\rab$ the following
q-dimensional representation of \tHooft :
\eqn\representation{ \eqalign{ (a)_{\alpha \beta}& = e^{2\pi i\vartheta_a}
e^{-2 \pi i p/q \alpha}
\delta_{\alpha,\beta} \cr (b)_{\alpha \beta}& = e^{2\pi i\vartheta_b}
\delta_{\alpha,\beta+1},
\;\;\;\alpha,\beta = 0,1,...,q-1.\cr}}

\noindent It is easy to check that the ``characteristics''
$(\vartheta_a ,\vartheta_b)$ modulo $(m/q,n/q)$ (with $m,n$
relative integers) label all the inequivalent
unitary irreducible representations of \tHooft . The space
of classes of inequivalent (irreducible, unitary) representations of
\tHooft\ is  therefore isomorphic to a 2-dimensional torus
$T_{p/q}$.

It is has been  stated \elitzur -\bos\ that for the modular
group to act on the Hilbert space of holomorphic wave functions \transition\
one needs both $pq$ {\it even} and the
characteristics $\vartheta_{a,b}\equiv 0$ modulo $1/q$.
Since this is not quite correct, let us pause to discuss the issue
of modular invariance in some detail. (See also \poly -\jengo .)
Let us denote by $s,t,c$ the following {\it external} automorphisms
of the `t Hooft algebra \tHooft:
\eqn\auto{\eqalign{s :& \cases{a\rightarrow b^{-1}\cr b\rightarrow a\cr}
\cr
t :& \cases{a\rightarrow a\cr b\rightarrow e^{-i\pi p/q}ab\cr}\cr
c :&\cases{a\rightarrow a^{-1}\cr b\rightarrow b^{-1}.\cr}\cr}}
\noindent One can to verify that $s,t,c$ satisfy the modular
group relations, $s^2 = c$ and $(st)^3=1$ and that the ``conjugation''
operator $c$ commutes with the modular group generators, $sc=cs$, $tc=ct$.
The automorphisms $s,t,c$ map representations of \tHooft\ onto
generically inequivalent representations; therefore they induce
a non-trivial action on the torus $T_{p/q}$, the space  of classes of
inequivalent (irreducible, unitary) representations
of the `t Hooft algebra \tHooft .
This action, however, is {\it not} the ``standard'' action of the modular group
on the 2-dimensional torus, which is linear and homogenous
in the coordinates $(q\vartheta_a,q\vartheta_b)$.
Denoting by $s_{*} ,t_{*} ,c_{*}$ the action of $s,t,c$ induced
on $T_{p/q}$ one can
explicitly calculate from \representation\ that $t_{*}$ has
an {\it inhomogenous} term:
\eqn\nonstandard{\eqalign {c_{*}:&(q\vartheta_a,q\vartheta_b)\rightarrow
(-q\vartheta_a, -q\vartheta_b)\cr
s_{*}:&(q\vartheta_a, q\vartheta_b)\rightarrow
(-q\vartheta_b, q\vartheta_b)\cr
t_{*}:&(q\vartheta_a, q\vartheta_b)\rightarrow
(q\vartheta_a, q\vartheta_a +q\vartheta_b +pq/2),\cr}}
\noindent where $q\vartheta_{a,b}$ are real numbers modulo integers.
The vector space of a representation $\rab$, belonging in an
equivalence class which is invariant under
$s_{*} ,t_{*} ,c_{*}$, carries a (unitary) representation of the modular group
whose generators we will denote by $S,T$ and $C$.
Such a representation $\rab$ defines through \transition\ a vector space
of holomorphic wave functions $\wf$ which
supports a (unitary) representation of the modular
group. The generators of this modular representation will be indicated
below by $U(s)$, $U(t)$ and $U(c)$.
{}From \nonstandard\ it follows that {\it if} $pq$ is even
the only (up to equivalence) modular invariant quantization
corresponds to the (equivalence class of the) $\roo$ representation
of the `t Hooft algebra \tHooft , a fact already recognized in
the earlier literature on Chern-Simons theory \elitzur ,\bos .
However eq.\nonstandard\ also implies that  modular invariance can be
mantained for $pq$ odd as well
by choosing  a representation of the `t Hooft algebra
in the equivalence class of $\rhalf$. This was first
realized in \poly\ in the
context of the abelian Chern-Simons theory.
We will show in the following that
in the non-abelian theory the choice $k=p/2$ with $pq=p$ {\it odd}
(disregarded in \elitzur ,\bos\ on modular invariance grounds)
does actually lead to the characters forming the $D_{p+1}$ series
of modular invariants for ${\hat A}_1$ current algebra \ref\cappelli
{A. Cappelli, C. Itzykson
and B. Zuber, \np {\bf B280} (1987) 445.}. Since these conformal
models are well-defined on Riemann surfaces of arbitrary topology
it is likely that a modular invariant quantization of Chern-Simons
theory with $k$ integer and odd, extending to all genuses the quantization
that we will exhibit here for the torus topology, does exist.

In geometric quantization, in order to implement canonical transformations
which do not leave the polarization invariant (such as modular
transformations), the wave functions $\Psi_{\tau}(z)$ are also
regarded as dependent on the polarization $\tau \in \Pol$. The $\tau$
dependence is determined by the requirement that
quantum Hilbert spaces ${\cal H}_{\tau}$
relative to different $\tau$'s
be unitarily equivalent with respect to the hermitian forms
\eqn\product{\left(\Psi_\tau^{(1)}, \Psi_\tau^{(2)}\right)=
\int_{\Phc}dz\,d{\bar z}\;\tau_2^{-1/2}
e^{-{k\pi\over 4\tau_2}(z-{\bar z})^2}
\Psi_{\tau}^{(1)}({\bar z})^{*}\Psi_{\tau}^{(2)}( z)}
\noindent associated to the K\"ahler structure \form .
This implies that the wave functions $\Psi_{\tau}(z)$ should
be {\it parallel} with respect to a flat, unitary connection
on the vector bundle with base $\Pol$ and fibers ${\cal H}_{\tau}$
\axelrod . One has now to distinguish the cases when
$pq$ is even or odd. For $pq$ even, an orthonormal $p$-dimensional
basis for the $q$-components
parallel wave functions $\wf$ of the quantization of $\Phcs$ is given by :
\eqn\basis{ (\wfN)^{\alpha} \equiv \wfaN =
\theta_{ qN + p\alpha,pq/2} (\tau; z/q), \;\;\;
N = 0,1,...,p-1,}
\noindent where the $\theta_{n,m}(\tau; z)$ ($n$ integer modulo
$2m$) are level $m$ $SU(2)$ theta functions
\ref\kacalgebras {V.G. Kac, ``Infinite Dimensional Lie Algebras'',
Cambridge 1985.}:
$$\theta_{n,m}(\tau;z)\equiv \sum_{j\in Z}e^{2\pi im\tau (j+{n\over 2m})^2
+2\pi imz(j+{n\over 2m})}.$$
For $pq$ odd, we have seen that modular invariance
requires the representation of the `t Hooft algebra \tHooft\ to be
(equivalent to) $\rhalf$. With this choice, an orthogonal $p$-dimensional
basis of parallel holomorphic wave functions is:
\eqn\obasis {\eqalign{(\wfN)^{\alpha} &=
(-1)^{qN+p\alpha} \sum_{j\in Z}e^{i\pi pq\tau (j +N/p +q/\alpha +1/2)^2 +
i\pi p (z-q)(j +N/p +q/\alpha +1/2)}\cr
&=\theta_{q(2N+p)+2p\alpha,2pq}(\tau;z/2q) -
\theta_{q(2N-p)+2p\alpha,2pq}(\tau; z/2q),\cr
&-p/2< N < p/2.\cr}}

Among classical canonical transformations, reflections ${\hat c}$ around the
singular points in $\Ns$
\eqn\reflections{{\hat c} : z \rightarrow -z.}
\noindent are of special interest for our purposes.
${\hat c}$ will be implemented on the Hilbert space of wave functions $\wf$
by a unitary operator $U(c)$:
\eqn\weyl{ U(c) :\wf \rightarrow C\Psi(-z),}
\noindent where $C$ is a $q\times q$ unitary matrix acting on the ``internal''
indices $\alpha$, which implements the automorphism
$c$ defined in \auto\ on the vector space of $c_{*}$-invariant
representations of \tHooft :
\eqn\cauto{ C a^m b^n = a^{-m} b^{-n} C .}
\noindent For $a,b$ in both the representation $\roo$ (when $pq$ is even)
and $\rhalf$ (when $pq$ odd) the solution of \cauto\ is:
\eqn\cautosolution{ (C)_{\alpha, \beta} = \delta_{\alpha, -\beta}.}

\noindent The Hilbert spaces $\Hc$ spanned by the sections \basis\ and
\obasis\
split under the action of $U(c)$ into ``even'' and
``odd'' subspaces $\Hcpm$. For $pq$ even
an orthogonal parallel basis of $\Hcpm$ is
\eqn\pmbasis{(\Psi_N^{\pm})^{\alpha}\equiv \psi^{\alpha,(\pm)}_N (\tau; z)
= \theta_ {qN + p\alpha ,pq/2 }(\tau; z/q) \pm
\theta_{-qN + p\alpha ,pq/2}(\tau; z/q) ,}

\noindent while for $pq$ odd one has:
\eqn\pmobasis{\eqalign{(\Psi_N^{\pm})^{\alpha}=
&(\theta_ {q(2N+p)+2p\alpha ,2pq}(\tau; z/2q)\pm
\theta_{-q(2N+p)+ 2p\alpha ,2pq}(\tau; z/2q))\mp \cr
&(\theta_ {q(p-2N)+2p\alpha ,2pq}(\tau; z/2q)\pm
\theta_{-q(p-2N)+ 2p\alpha ,2pq}(\tau; z/2q)).\cr}}

Were we simply trying to quantize $\Phcs$, we would keep both the
even and the odd sector since canonical transformations \reflections\
in the $\Phc$ sector do {\it not} correspond to gauge
transformations of the original Chern-Simons \G\ theory.
However, we really want  to quantize
the union $\Phc \cup \Phn $. There are no ``rigorous'' ways to
quantize a phase space consisting of different branches
with a non-zero intersection. Phase spaces of this sort have
appeared in the context of 2-dimensional gravity in
\ref\witmatrix{E. Witten, IAS preprint, SLAC-PUB-IAP-HEP-91/26.}.
It seems reasonable to think of a wave function on the union
$\Phc \cup \Phn$ as a pair $(\psi_1,\psi_2)$ of wave functions,
with $\psi_1 \in \Hc$ and $\psi_2 \in \Hn$, ``agreeing''
in some sense on the intersection $\Ns$. Our proposal is that
$\psi_1$ and $\psi_2$ should have the same behaviour around the
points in $\Ns$. Since $\psi_1$ and $\psi_2$ are represented
by holomorphic functions, this implies that the pair $(\psi_1,
\psi_2)$ should be determined uniquely by $\psi_1$ and
that most of the states $\psi_2$ in the infinite-dimensional
$\Hn$ should be discarded. Moreover, the B\"ohm-Aharonov phase
$e^{2\pi i\vartheta}$ in the $\Phn$ branch should coincide
with the analogous quantity $e^{2\pi ip/q}$ in the $\Phc$ sector.
However, all states $\psi_2$ in $\Hn$ have the same behaviour under reflections
${\hat c}$ around singular points, since ${\hat c}$ corresponds to a gauge
transformation of the Chern-Simons theory in the $\Phn$ branch.
This should put a restriction on the states $\psi_1$, which, in
order to ``agree'' with $\psi_2$, should also have definite parity
under ${\hat c}$. In conclusion the (only) r\^ole of $\Phn$
should be ``transmitting'' to $\Phc$
the definite ${\hat c}$-parity projection.
The phase space $\Phc \cup \Phn$ admits therefore two inequivalent
quantizations, with Hilbert spaces isomorphic to $\Hcpm$ .
A similar ambiguity is present in the $SU(2)$ case \imbimbo ,
but it is the ``odd'' quantization which is related to 2-dimensional
conformal field theories for generic $k$. In fact, only the ``odd''
projection gives positive integer fusion rules for $k$ generic, suggesting
that this is the quantization of the Chern-Simons theory on the torus
which generalizes, in some appropriate sense, to higher genus space
manifolds \axelrod .
In our case as well, ``odd'' quantization gives positive, integer fusion
rules for generic $k$, as we will shortly see, though
we do not yet know its 2-dimensional interpretation.

Wave functions in $\Hcm$ are related to the Kac-Wakimoto characters
\vgkac\ of irreducible, modular invariant representations of
\G\ current algebra with fractional central charge $m\equiv t/u$
($t,u$ coprime integer relative numbers, $u$ positive) satisfying the
{\it admissibility} condition
\eqn\admissible{2u+t-2\geq 0. }
\noindent The Kac-Wakimoto characters
are defined as follows:
\eqn\characters{ \chi_{j(N^{'}, \alpha^{'});m} (z,\tau)=
tr_{{\cal H}_{j,m}} e^{ 2 \pi i \tau L_o + 2 \pi i z J^3_o}, }

\noindent where ${\cal H}_{j,m}$ is the highest weight irreducible
representation of $SL(2,R)$ current algebra with level $m$ and
spin $j$. $\;j =j(N^{'}, \alpha^{'})$  ranges over the following set:
\eqn\spin{ j = 1/2 ( N^{'} - \alpha^{'} (m+2) ), \;\;\;
N^{'} = 1, 2,...,2u + t-1, \;\;\; \alpha^{'} = 0,1,...u-1.}

\noindent In order to exhibit the explicit relation between
Kac-Wakimoto characters and Chern-Simons wave functions $\wf$
one has to distinguish the cases when:

\noindent (i) $p$ is even and $q$ is odd, so that $p=2s$ and $q=r$;

\noindent (ii) $p$ is odd and $q$ is even, so that $p=s$ and $r=2q$ is a
multiple of 4;

\noindent (iii) both $p$ and $q$ are odd, so that $p=s$ and  $r=2q\equiv 2$
mod 4.

In case (i) the ``odd'' orthogonal wave functions in \pmbasis\
can be written in terms of the Kac-Wakimoto characters $\chi_{j;m}$
of level $m$ given by :
\eqn\level{m+2=k,}
\noindent i.e. $u=q=r$ and $p/2=s=2u+t$. The explicit relation is:
\eqn\kac{ {\psi^ {\alpha,(-)}_N (\tau; z) \over \Pi (\tau; z)} =
\cases { \chi_{j (N, 2\alpha);m}(\tau;z) & if
$\alpha \in \{0,1,...,(r-1)/2\}$ \cr
\chi_{j (s-N, 2\alpha - r);m}(\tau;z) & if
$\alpha \in \{ (r+1)/2, ..., r-1\}$, \cr} }

\noindent where $\Pi (\tau ;z)$ is the Kac-Wakimoto denominator:
\eqn\denominator{ \Pi (\tau ;z) = \theta_{1,2} (\tau ,z) -
\theta_{-1,2} (\tau, z).}
\noindent $\Pi (\tau ;z)$ is holomorphic and non-vanishing on $\Phcs$.
Therefore, the wave functions $\Psi^{(-)}_N (\tau; z)$  and the wave functions
$$ \Psi^\prime_N (\tau;z)=
{\Psi^{(-)}_N (\tau; z) \over \Pi ( \tau; z)}$$
\noindent  appearing in \kac, describe equivalent wave functions
on $\Phcs$, related to each others by a K\"ahler
transformation
\eqn\trivialization{K \rightarrow K + f(z) + f^* ({\bar z}),}
\noindent with $f(z) = - ln (\Pi (\tau ;z))$. Eq.\trivialization\ introduces
in the scalar product \product\ a factor which
has been interpreted as the jacobian of a certain change of integration
variables in the path integral formulation of the Chern-Simons theory
\elitzur . Eq.\kac\ implies that
to each quantum state emerging out of the quantization of the
\G\ Chern-Simons theory with fractional $2k$
there corresponds a {\it multiplet} of
$SL(2,R)$ current algebra characters, rather than one single character
as in the $SU(2)$ case when $2k$ is integer.
One does not expect, therefore, that the hypothetical 2-dimensional
theory underlying 3-dimensional \G\ Chern-Simons theory has
full \G\ current algebra symmetry. One might speculate that such a
theory could be obtained from some coset construction of
\G\ current algebra,
though not a standard coset as will become apparent from the analysis
of the modular transformation properties which will be studied in
the following section.

If (ii) holds, ``odd'' wavefunctions in \pmbasis\ are
expressible in terms of Kac-Wakimoto characters of level
\eqn\levelbis{m+2=4k}
\noindent (i.e. $u=q/2=r/4$ and $p=s=2u+t$),
as it follows from the identities:
\eqn\kacbis{{\psi^{\alpha,(-)}_N (\tau; z) \over \Pi (\tau; z)} =
\cases {\chi_{j (2N,\alpha);m}(\tau;z) & if $\alpha \in \{0,1,...,r/4-1\}$\cr
\chi_{j (s-2N,\alpha - r/4);m}(\tau;z) & if $\alpha \in \{r/4, ..., r/2-1\}$.
\cr }}

Finally, if (iii) is true, the level $m$ of the \G\ current algebra
is still given by eq.\levelbis , but $u=q=r/2$ and $2u+t=2p=2s$.
The relation between wave functions and characters becomes:
\eqn\kactris{ {\psi^ {\alpha,(-)}_N (\tau; z) \over \Pi (\tau;z)} =
\chi_{j(p+2N,\alpha);m}(\tau;z/2) +\chi_{j(p-2N,\alpha);m}(\tau;z/2).}

In all cases (i)-(iii), the Kac-Wakimto admissibility condition is
equivalent to the statement that ``odd'' projection $\Hcm$ be
non-empty (i.e. $s\geq 2$).

When the Chern-Simons charge $k$ is an integer, i.e. $r=1=q$ and
$p=2k=2s$ (case (i)), eq.\kac\ reduces to the well-established
\elitzur -\bos\ identification between Chern-Simons wave functions and
integrable ${\hat A}_1$ Kac-Weyl characters of level $m=k-2$
forming the diagonal modular invariant $A_{k-1}$ series
of the classification of Cappelli et al.\cappelli .

Wave functions are one-dimensional vectors of holomorphic
functions also if $k=p/2$ is {\it half-integer} (and $p=s$ odd, $q=r/2=1$).
This case belongs in (iii), therefore the level $m=2(p-1)$ of the
current algebra is a multiple of 4. Eq.\kactris\ becomes:
\eqn\dseries{\psi^\prime_N (\tau; z)=
{\psi^{(-)}_N (\tau; z) \over \Pi (\tau;z)} =
\chi_{p+2N;2(p-1)}(\tau;z/2) +\chi_{p-2N;2(p-1)}(\tau;z/2),}
\noindent from which one sees that wave functions are
precisely those linear combinations of ${\hat A}_1$
characters $\chi_{n;2(p-1)}$ of level $2(p-1)$ which
form the $D_{p+1}$-series of \cappelli .

\beginsection 3. Modular Transformations

Let us use ${\hat s}$ and ${\hat t}$ to denote the canonical transformations
of the classical phase space $\Phcs$ which generate the modular group
$SL(2,Z)$ of the torus
\eqn\modular{\eqalign{{\hat s}:&\;\; (\tau, z) \rightarrow (-1/\tau , z/\tau )
\cr
{\hat t}:&\;\; (\tau, z) \rightarrow ( \tau + 1, z), \cr} }
\noindent and satisfy the relations:
\eqn\Group{ {\hat s}^2 ={\hat c}, \;\;\; ({\hat s}{\hat t})^3 = 1.}
\noindent ${\hat s}$ and ${\hat t}$ will be represented on the space of
the multi-components wave functions $(\wf)^\alpha
\equiv \psi^{\alpha}(\tau; z)$ by means of unitary operators
$U(s)$ and $U(t)$:
\eqn\action{\eqalign{ U(s) &: \Psi (\tau; z) \rightarrow (S^{-1}\Psi )(
-1/\tau;
z/\tau) \cr
U(t) &: \Psi (\tau; z) \rightarrow ( T^{-1}\Psi) ( \tau + 1; z), \cr } }

\noindent where $S \equiv (S)_{\alpha\beta}$ and $T \equiv (T)_{\alpha\beta}$
are unitary $q \times q$ matrices acting on  the ``internal'' indices and
implenting the modular transformations \auto\ on the representation space
the `t Hooft algebra \tHooft . Choosing the $\roo$ representation of
\tHooft\ when $pq$ is even and $\rhalf$ when $pq$ is odd, one
obtains the following expressions for matrices $T$ and $S$:

\eqn\TSeven{\eqalign{ (T^{(p;q)})_{\alpha\beta} &= \delta_{\alpha,\beta}
(-1)^{pq\alpha}e^{ 2\pi i {p \over 2q} \alpha^2 - 2\pi i \theta (p;q)/3 } \cr
(S^{(p;q)})_{\alpha\beta} &= {1\over \sqrt{q}}
e^{2\pi i {p \over q} \alpha\beta}, \cr
\alpha, \beta &= 0,1,...,q-1.\cr}}

\noindent The phase $\theta (p;q)$ in \TSeven\ is determined from
the $SL(2,Z)$ relation $(ST)^3 =1$, which gives:

\eqn\tetapq{ e^{2\pi i \theta(p;q)} = {1\over \sqrt{q}}\sum_{n=0}^{q-1}
(-1)^{pqn}e^{2\pi i
{p \over 2q} n^2}. }

\noindent When $p=1$ (and $q$ is even)
this is the celebrated Gauss sum \ref\gauss
{See for example, Estermann, J. London Math. Soc. {\bf 20} (1945) 66.}, and
it is well-known that $\theta(1;2K) \equiv1/8$ mod $1$, agreeing with the fact
that the conformal central charge of a free compactified 2-dimensional
scalar field is one. In fact, ${\bf R}^{(1;2K)}$ is the representation
of the modular group associated to the conformal blocks of a
2-dimensional scalar field  compactified on a circle of radius
$R^2 = 2m/n$ with $m$ and $n$ integers and $K=mn$.
${\bf R}^{(1;2K)}$ is also the representation of the modular group that
one obtains upon quantization of the abelian Chern-Simons theory
on a torus \elitzur -\bos . For $p\not=1$, the sum in \tetapq\
is a generalized Gauss sum which has not yet appeared in
conformal field theory and which we calculate
in the Appendix. Some properties of $\theta(p;q)$
follow immediately from its definition \tetapq :
\eqn\residue{ \eqalign{\theta (p + 2q;q) &\equiv \theta(p;q) \;\; mod \;1\cr
\theta (p';q)& \equiv \theta (p;q) \;\; mod \;1, if\;\;
p'\equiv p n^2 \;\;mod\; 2q \cr} }
\noindent for $n$ integer. The explicit
formula for $\theta (p;q)$ derived in the Appendix implies that
\eqn\otto{ e^{8\pi i\theta (p;q)} = -1,}
\noindent i.e., that the allowed values for $\theta (p;q)$ are $\pm 1/8$
and $\pm 3/8$ (mod $1$).
\bigskip
The representations of the modular group acting on the quantum Hilbert
spaces $\Hc$ and $\HCS$ can now be derived from the modular properties
of the theta functions in \basis ,\obasis ,\pmbasis ,\pmobasis\ and from
the representation \TSeven\ acting on the ``internal'' indices.
$\Hc$ carries the $p-$dimensional
representation $\Rqp$ ``dual'' to the representation
$\Rpq$ defined in \TSeven :
\eqn\reprqp{\eqalign{T_{N,M}^{(q;p)} &= (-1)^{Npq}
e^{2\pi i{q \over 2p}N^2 - 2\pi i \theta (q;p)/3} \delta_{N,M}  \cr
S_{N,M}^{(q;p)} &= {1\over \sqrt{p}}e^{2\pi i {q \over p}N M} \;\;\;
N,M = 0,1,...,p-1. \cr } }

\noindent This representation is equivalent to the representation
of the modular group obtained in \poly ,\jengo\ by quantizing
{\it abelian} Chern-Simons theory with fractional coupling constant.
For $q\not\equiv n^2$ mod $2p$ its interpretation
in terms of 2-dimensional conformal field theories is still obscure.
We concentrate, however, on the representations carried by
$\HCS \equiv \Hc^{\pm}$. Note that $(\Sqp )^2 = C$,
with $ (C)_{N,M} = \delta_{N, -M}$ being the ``charge
conjugation'' matrix. Since $C$ commutes with the matrices in \reprqp ,
$\Rqp$ decomposes into two representations ${\bf R}^{(q;p)}_{\pm}$,
even and odd under $C$:
\eqn\reprpm{ \Rqp = \Rqp_{+} \oplus \Rqp_{-},}
\noindent where $\Rqp_{\pm}$ is $p/2 \pm 1$ dimensional if $p$
is even (i.e., if $r$ is odd) and $(p\pm 1)/2$-dimensional
if $p$ is odd (i.e., if $r$ is even). Since $(S^{(q;p)}_{-})^2 =
- {\bf 1}$, it is convenient to define a ${\bf \tilde R}^{(q;p)}_{-}$
by ${\tilde S}_{-}^{(q;p)}= -i S^{(q;p)}_{-}$ and
${\tilde T}_{-}^{(q;p)}= i T^{(q;p)}_{-}$
such that the charge conjugation matrix is equal to the
identity.
When $q=1$ the ``odd'' representation
${\bf \tilde R}^{(1;p)}_{-}$
is the one associated to modular invariants of ${\hat A}_1$ current algebra
(to the diagonal $A_{p/2-1}$ series of level $p/2-2$ if $p$ is even,
to the $D_{p+1}$ series of level $2p-2$ if $p$ is odd.).
The fusion rules associated to the ``even'' representation are
not positive and integer-valued for generic $p$, suggesting
that the ``even'' quantization  does not extend to Chern-Simons
theories defined on higher genus surfaces \axelrod ,\imbimbo .
The same remains true for generic $q$. This motivates the
choice $\HCS = \Hc^{-}$ carrying the modular representation
$\RCS \equiv {\tilde \Rqp}_{-}$ which has the following
explicit matrix representation:

\eqn\reprcs{\eqalign{T^{CS}_{N,M} &= i (-1)^{Npq}
e^{2\pi i{q \over 2p}N^2 - 2\pi i \theta (q;p)/3} \delta_{N,M}  \cr
S^{CS}_{N,M} &= {1\over \sqrt{p}}\sin 2\pi i {q \over p}N M \;\;\;
N,M =1,...,[(p-1)/2], \cr } }
where $[x]$ is the largest integer $\leq x$.
Therefore, the central charge $c$ and the conformal dimensions
$h_N$ of a hypothetical 2-dimensional conformal field theory
underlying the
3-dimensional \G\ Chern-Simons theory should satisfy the
equation
\eqn\charge{ h_N - c/24 = N^2/{4k} - \theta (q;p)/3 + 1/4 \;\;mod\; 1.}

\noindent If such theory were unitary and had a unique
identity operator corresponding to the conformal block
labelled by ${\bar N} \in \{1,2,...,[(p-1)/2]\}$, one would
have:

\eqn\spectrum{ \eqalign{ c &= 2 - 12 {\bar N}^2 q/p +
8\theta (q;p)\;\;\; mod\; 8\cr
h_N &= {q\over 2p}(N^2 - {\bar N}^2)\;\; mod\; 1. \cr}}

%\noindent Note that for $q=1$ (and
%${\bar N}= 1$), the
%spectrum and central charge in \spectrum\ are those of ${\hat A}_1$
%current algebra with level $l= p/2-2$ and spin $j= (N-1)/2$.

Since eqs.\kac\ and \kacbis\ express the multi-component Chern-Simons
wave functions in terms of Kac-Wakimoto $SL(2,R)$ characters
with level $m$ given by Eqs.\level\ and \levelbis , the
modular representation $\RCS$ is related, when $pq$ is even,
to the Kac-Wakimoto
representation ${\bf R}_{KW}^{(m)}$ through the equation:
\eqn\wak{ {\bf R}_{KW}^{(m)}= {\bf R}^{(p;q)} \otimes \RCS.}
When $pq$ is odd, a similar equation holds with the l.h.s. given
by the modular representation acting on the Kac-Wakimoto
characters which appear in \kactris\ and which define
a generalization of the $D$-series to the fractional level
case.
Eq. \wak\ encodes the relationship
between Chern-Simons theories and Wess-Zumino-Witten models
when $2k$ is fractional. For integer
$2k$ (i.e., for $q=1$), the left factor on the r.h.s. of \wak\
is trivial, and one obtains the well-established
correspondence between Chern-Simons states and
current algebra blocks. For fractional $2k$, Eq. \wak\ can be
phrased by saying that the
2-dimensional theory underlying \G\ Chern-Simons theory
is the ``quotient'' of \G\ current algebra by
some yet unknown generalization of the gaussian model
whose modular properties are given by $\Rpq$.

It was discovered
in \ref\mp{S. Mukhi and S. Panda, \np\ {\bf B338} (1990) 263.}
that the Kac-Wakimoto characters are related
by means of a certain projection to the
Rocha-Caridi characters of the $c<1$ conformal discrete
series.
This suggests that the modular representation
$\RCS$ in \reprcs\ has something to do with the
representation ${\bf R}^{(r;s)}_{Vir}$ relative
to the $(r,s)$ minimal model of Belavin-Polyakov-Zamolodchikov.
This in fact turns out to be the case
and one can establish, when $pq$ is even,  the following equation:
\eqn\vir{ {\bf R}^{(r;s)}_{Vir} = {\bf R}^{(r/4s)}_{CS} \otimes
{\bf R}^{(s/r)}_{CS},}

\noindent where $r$ must be chosen {\it odd}. (This is always
possible since $r$ and $s$ are coprime integers:
however, the r.h.s. of the Eq.\vir\ is not symmetric
under the interchange of $s$ and $r$ if one of them is even.
The equation as written is {\it not} valid for $r$ even.)
In order to understand how \vir\ comes about, let us consider the
abelian Chern-Simons theory with {\it even} integer charge $K=pq$ whose
algebra of observables ${\it O}_K$ is generated by the holonomies $A$ and $B$
around the non-trivial cycles of the torus \elitzur ,\bos :
\eqn\observables{ AB = BA e^{2\pi i/K}.}
\noindent The quantum Hilbert space is K-dimensional and
spanned by $SU(2)$ theta functions $\theta_{\lambda,K/2}$
(with $\lambda \in Z_{K}$) of level $K/2$. It carries the  representation
${\bf R}^{(1;K)}$ of the modular group. Now the crucial fact
is that
\eqn\decomposition{{\it O}_K \approx {\it O}_{p/q} \times {\it O}_{q/p},}
\noindent where ${\it O}_{p/q}$ (${\it O}_{q/p}$) is
a `t Hooft algebra  defined as in \observables , with
${\tilde {\it A}}\equiv A^p , \; {\tilde {\it B}}\equiv B^p$
(${\tilde {\it A}}\equiv A^q ,\;{\tilde {\it B}}\equiv B^q$)
and ${\it O}_{p/q}$,${\it O}_{q/p}$ {\it commute} among themselves.
Therefore ${\bf R}^{(1,K)}$ factorizes:
\eqn\factor{{\bf R}^{(1,K)} = \Rpq \otimes \Rqp .}

\noindent The `t Hooft algebra ${\it O}_K$ is invariant under
a conjugation $C_K$, $C_K :A \rightarrow A^{-1}, B\rightarrow B^{-1}$,
which is in the commutant of the representation ${\bf R}^{(1,K)}$ and,
therefore, can represented by a diagonal matrix in the representation
space of ${\bf R}^{(1,K)}$. In the holomorphic representation of
${\it O}_K$ spanned by the theta-functions $\theta_{\lambda,K/2}$,
the operator $C_K$ acts as follows:
\eqn\conjugation{C_K :\lambda \rightarrow -\lambda .}
\noindent ${\bf R}^{(1,K)}$ decomposes into
two modular representations, ``even'' and ``odd'' under $C_K$ and
the ``odd'' representation is associated to non-abelian current
algebra (of level K-2), as mentioned above. The new fact which occurs
when $K=pq$ is not prime is that, because of the decomposition
\decomposition , the group $W_K$ of conjugations of ${\it O}_K$
is enlarged to a {\it four} element group $Z_2\otimes Z_2$, generated
by the conjugations $C_{p/q}$ and $C_{q/p}$ of the algebras
${\it O}_{p/q}$ and ${\it O}_{q/p}$, with $C_K = C_{p/q}C_{q/p}=
C_{q/p}C_{p/q}$. Decomposing $\lambda \in Z_K\approx Z_p \otimes Z_q$
in terms of $M\in Z_q$ and $N\in Z_p$,
$\lambda \equiv Mp - Nq$, the action of the conjugation operators is:
\eqn\action{\eqalign{C_{p/q} &: \lambda \rightarrow {\bar \lambda}\equiv
Mp + Nq\cr C_{q/p} &: \lambda \rightarrow -{\bar \lambda} .\cr }}
The existence of extra conjugation operators opens
the possibility of considering several kinds of
projections of the representation ${\bf R}^{(1,K)}$
according to the values of $C_{p/q}$ and $C_{q/p}$.
Projecting to the {\it odd} sector of a single conjugation
operator (let us say $C_{q/p}$),
one obtains the modular representation of the Kac-Wakimoto
of level $m$ given by Eqs.\level ,\levelbis , as
apparent from \wak . Considering instead the subrepresentation
which is completely anti-symmetric with respect to the whole
conjugation group $W_K$ of ${\it O}_K$, one obtains the representation
of the modular group relative to the $(r,s)$ minimal models,
where $r,s$ are defined through $K/2 =pq/2 \equiv rs$.
In fact, the completely anti-symmetric
holomorphic wave functions are:
\eqn\anti{\chi_{\lambda}(z;\tau) = \theta_{\lambda ,K/2} -
\theta_{{\bar \lambda },K/2}
+\theta_{-\lambda ,K/2} -\theta_{-{\bar \lambda},K/2}, }
\noindent with $\lambda \in \Gamma$, where $\Gamma \subset Z_K$ is any
fundamental set for the action of $W_K$ on $Z_K$. $\chi_{\lambda}(0;\tau)$
are nothing but the numerators of the Rocha-Caridi characters
of the completely degenerate representations of the $c<1$
$(r,s)$ minimal models. Since $\RCS$ is the odd projection
(with respect to $C_{q/p}$) of $\Rqp$, this establishes Eq.\vir .
%Note that the modular
%transformation rules of the Rocha-Caridi characters are not
%usually presented in the factorized form \vir\ because of the
%familiar choice of the fundamental set $\Gamma$ is
%$\Gamma_{o} =\{\lambda = nr-ms: 1\leq n\leq s-1 \;1\leq m\leq r-1,
%\lambda >0\}$. However, $\Gamma_{o}$ is equivalent to the domain
%$\Gamma_{1} \equiv \{\lambda = Nq-Mp: 1\leq N\leq p/2-1 \;1\leq M\leq
%(q-1)/2,\}$, taking $p=2s$ even and $q=r$ odd, for which
%the factorization property \vir\ is manifest.

Loosely speaking, Eq.\vir\ tells us that the hypothetical
2-dimensional conformal field theory corresponding to \G\ Chern-Simons
topological theory with fractional charge $k=s/r$ can be regarded as
the ``{\it square root}'' of conformal $(r,s)$ minimal models. More precisely,
Eq.\vir\ states that Chern-Simons theory with gauge group
$SL(2,R) \times SL(2,R)$ and charges $(k, 1/{4k})$ with $k=s/r$
rational has the modular properties of the $(r,s)$ minimal
model if $r$ is odd, and of the $(r/4,s)$ minimal model if $r$ is even
(in which case it must be a multiple of 4).

%It is intriguing that the integers $(r,s)$ enter the r.h.s. of
%Eq.\vir\ asymmetrically, though the l.h.s.
%is explicitely symmetric. An analogous phenomenon occurs when
%coupling  $(r,s)$ conformal minimal models to 2-dimensional
%gravity \ref\dvv{R. Dijgraaf, E. Verlinde and H. Verlinde,
%\np  {\bf B348} (1991) 435.}, a circumstance which may support the conjecture
%that \G\ Chern-Simons theory is intimately related to 2-dimensional
%quantum gravity \hverlinde .

The fact that the states of \G\ Chern-Simons theory are labeled
by one of two indices appearing in minimal models formulas,
seems to suggest that they correspond to
Virasoro representations at the ``boundary'' of the Kac table.
The conformal dimensions $\Delta_{N,0}$ with $N=1,...,s-1$
of the degenerate ``boundary'' representations of the minimal
$(r,s)$ model satisfy the equation
\eqn\boundary{ \Delta_{N,0} - c/24 = rN^2/4s - 1/24,}
\noindent which looks almost the same as the corresponding
equation \charge\ for the Chern-Simons theory, were it not for
the phase $\theta (q;p)/3$ not equal, for generic $(r,s)$,
to 1/24 . When considering the tensor product of two Chern-Simons
representations, as in \vir , the two phases $\theta (q;p)/3$
and $\theta (p;q)/3$ add up to produce the 1/24 required
by the Kac formula. Actually, degenerate Virarsoro
representations at the boundary of the Kac table are
not closed under modular transformations, so that the disagreement
between \charge\ and \boundary\ is not truly surprising. However,
considering also the apparently  important
role that boundary representations play in the context
of string theory in $c<1$ conformal backgrounds, we feel that
the closeness between \boundary\ and \charge\
may nevertherless be significant.

The algebraic data needed to reconstruct a 2-dimensional rational
conformal field theory is not exhausted by the modular representation
of the conformal blocks of the genus one partition function.
The representations of the modular group associated with
conformal field theories, together with the braid matrices,
should satisfy a set of
rather restrictive identities, known as ``the polynomial equations''
\ref\moore{G. Moore and N. Seiberg, \pl\ {\bf B212} (1988) 451; \cmp\
{\bf 123} (1989) 177.}.
In the Chern-Simons framework, the derivation of the braid matrices
would require the solution of the theory on a space-manifold with
the topology of a sphere with punctures, a problem which we did not
address here. The polynomial equations imply however certain
necessary, but not sufficient, conditions for the representation of the
modular group of the conformal blocks of the identity operator on the torus.
The most celebrated among these conditions, due to E. Verlinde
\ref\verfusion{ E. Verlinde, \np\ {\bf B300} (1988) 360.},
requires that the numbers $N_{ijk}$, defined in terms of the
modular matrix ${\cal S}$ as
\eqn\fusion{ N_{ijk}= \sum_{n} {{\cal S}_{in} {\cal S}_{jn} {\cal S}_{kn}
\over {\cal S}_{0n} },}

\noindent be positive and integer, since they are interpreted as the
{\it fusion rules} of the conformal field theory. From the expression
for ${\cal S}_{CS}^{(s/r)}$, derivable from \reprqp -\reprcs ,
one obtains the same (positive and integer) fusion rules
as for the $SU(2)$ Wess-Zumino model of level $[{p-3\over 2}]$:
\eqn\rules{ N_{ijk}^{CS} = N_{ijk}^{SU(2)_{[{p-3\over 2}]}}.}
\noindent \rules\ can be thought of as the ``square-root''
of the Virasoro minimal models fusion rules, in agreement with \vir .

Modular invariance of 4-point correlation functions of primary
operators on the sphere gives rise to another necessary condition
for the $h_p$'s in \spectrum\ to be the spectrum of dimensions
of a conformal field theory \ref\vafa{C. Vafa, \pl\ {\bf B206} (1988) 421.}.
This condition requires that
\eqn\condition{\left(\alpha_i\alpha_j\alpha_k\alpha_l\right)^{N_{ijkl}}=
\prod_{r}\alpha_r^{N_{ijkl,r}},}
\noindent where $\alpha_i = \exp(2\pi i h_i)$ and
$$N_{ijkl,r} = N_{ijr}N_{klr} + N_{jkr}N_{ril} + N_{ikr}N_{rjl}.$$
\noindent One can verify that \condition\ is indeed satisfied by the
dimensions in \spectrum\ and the fusion rules in \rules .

The fact that the representations $\RCS$ of the modular group satisfy
both Verlinde and Vafa conditions is quite remarkable,
and makes $\RCS$ interesting from the point of view of conformal
field theories regardless of the Chern-Simons framework
which we adopted to derive them.
$\RCS$ is invariant for $q\rightarrow q+2p$ and representations
$\RCS$ and ${\bf R}_{CS}^{(s/r')}$ with the same $p$ for which
\eqn\equivalence{q^\prime \equiv q n^2 mod\;2p,}
with $n$ integer, are unitarily equivalent (as it follows from
\residue ). Thus, for each given $p$, as $q$ varies one obtains
a finite number of inequivalent modular representations $\RCS$
one for each equivalence class in $Z_{2p}/\sim$, where $\sim$ is
the equivalence relation in $Z_{2p}$ defined by \equivalence .

The equivalence class of $q=1$ corresponds to the usual modular
representations relative to ${\hat A}_1$ current algebra
with integer level.
The $\RCS$'s for low values of $s$ (and $q\not\sim 1$)
coincide with modular group
representations associated with Wess-Zumino-Witten models on
group manifolds other than $SU(2)$.
For example, for $s=3$, $\RCS$ is 2-dimensional and, according to
the value of $r$, coincides either with the representation of
the modular group associated to $SU(2)_1$ or with that associated
with $(E_7)_1$.
For $s=4$, $\RCS$ is 3-dimensional, and as $r$ runs over the
integers comprime with $s=4$ modulo $16=4s$,
one obtains the representations of the modular group of
$SO(2N+1)_1$ with $N=2,3,...,7$ and of the Ising model (for $r=3$).
But already for $s=5$, the representations $\RCS$ for $q\not\sim 1$,
do no longer appear to be equivalent to modular group representations
coming from current algebras.

Unlike in the compact case, the geometric quantization of \G\ Chern-Simons
theory does not provide the explicit functions of Teichm\"uller space
which transform according to the representation $\RCS$ of the modular
group and which are identifiable  with the Virasoro characters of an
underlying 2-dimensional conformal theory.
Actually, although conditions \fusion\ and \condition\
are both satisfied, this 2-dimensional ``object'', at least for generic $k$,
might belong to a class of quantum field theories
more general than the conformal family --- a class
of theories for which concepts like holomorphic blocks, modular invariance
and fusion rules  would be still be meaningful.
It is tempting to speculate that the 2-dimensional theories associated to
\G\ Chern-Simons theories are related to quantum deformations of \G\
Kac-Moody algebras with quantum parameter
$q= e^{4\pi ik}$, to which several concepts of conformal
field theories extend.

\bigskip
\beginsection ACKNOWLEDGEMENTS

I would like to acknowledge stimulating conversations with
M. Caselle, V.G. Kac, R. Jengo, A. Rosly, A Schwimmer  and M. Wakimoto.

\appendix{A}{}

The following theta function
\eqn\thetaeven{\theta (\tau) = \sum_{n\in Z} e^{i\pi n^2 \tau}}
is a holomorphic modular form of weight 1/2 for the subgroup
$\Gamma_{\theta}$ of $SL(2,Z)$ \kacalgebras :
\eqn\form{\theta (\alpha (\tau)) = e^{2\pi i \phi (\alpha)}
({c\tau +d\over i})^{1/2}\theta (\tau),}
where $\alpha \equiv \left (\matrix{a&b\cr c&d\cr}\right)\in
\Gamma_{\theta}$, with $c>0$, $\alpha (\tau) \equiv
{a\tau +b \over c\tau +d}$ and $\phi (\alpha)$ is a phase
defining the multiplier system of $\theta (\tau)$.
We will also need the weight 1/2 modular form
\eqn\thetaodd{ {\tilde \theta}(\tau) = \sum_{n\in Z}(-1)^n e^{i\pi n^2 \tau }}
whose multiplier system is $e^{2\pi i{\tilde \phi}(\alpha)} =
e^{2\pi i\phi (t\circ \alpha\circ t^{-1})}$, with  $\alpha \in t^{-1}
\Gamma_{\theta} t$. Let us consider the limit of \thetaeven\
and \thetaodd\ for
\eqn\cusp{\tau \rightarrow p/q +i\epsilon , \;\;\epsilon\rightarrow 0^{+}
\; q >0.}
{}From the definitions \thetaeven ,\thetaodd\ one derives the asymptotic
expressions:
\eqn\asymptotic{\eqalign{\theta(\tau )=
\sum_{n\in Z}e^{i\pi{p\over q}n^2}e^{-\pi n^2\epsilon}
&{\buildrel \tau \to p/q +i\epsilon \over \approx}
\sum_{n^2\leq {1\over \sqrt{\epsilon}}} e^{i\pi {p\over q}n^2}\cr
&\approx {1\over\sqrt{q\epsilon}}{1\over \sqrt{q}}
\sum_{n=0}^{n=q-1}e^{i\pi{p\over q}n^2}\cr}}
and
\eqn\asymptodd{{\tilde \theta} (\tau ) = \sum_{n\in Z} (-1)^n
e^{i\pi{p\over q} n^2} e^{-\pi n^2 \epsilon}
{\buildrel \tau \to p/q +i\epsilon \over \approx}
{1\over\sqrt{q\epsilon}}{1\over\sqrt{q}}
\sum_{n=0}^{n=q-1}(-1)^n e^{i\pi{p\over q} n^2}.}

When $pq$ is even one can find (not uniquely) an element $\alpha \in
\Gamma_{\theta}$, such that
\eqn\find{\alpha = \left (\matrix{a&b\cr q& -p\cr}\right).}
When $pq$ is odd a matrix $\alpha$ satisfying \find\ can be found in
the modular subgroup $t^{-1}\Gamma_{\theta} t$. In the limit \cusp\
one has
$$e^{2\pi i\alpha (\tau)}\to 0$$
\noindent with $\alpha$ satisfying \find , so that either
$$\theta (\alpha (\tau))\to 1$$
\noindent (for $pq$ even) or
$${\tilde \theta}(\alpha (\tau)) \to 1$$
\noindent (for $pq$ odd).
Therefore the modular properties \form\ imply the aymptotic expression
\eqn\newasym{\theta (\tau ){\buildrel \tau \to p/q +i\epsilon
\over \approx}{1\over\sqrt{q\epsilon}} e^{-2\pi i \phi (\alpha)},}
if $pq$ is even, and
\eqn\newasymodd{{\tilde \theta}(\tau ){\buildrel \tau \to p/q +i\epsilon
\over \approx}{1\over\sqrt{q\epsilon}}e^{-2\pi i {\tilde \phi}(\alpha)},}
if $pq$ is odd. Comparing with \asymptotic\ and \asymptodd\ one concludes
that
\eqn\phaseven{e^{2\pi i\theta(p;q)} = e^{-2\pi i\phi (\alpha)}}
if $pq$ is even, and
\eqn\phaseodd{e^{2\pi i\theta(p;q)} = e^{-2\pi i{\tilde \phi} (\alpha)}=
e^{-2\pi i\phi (t\circ \alpha \circ t^{-1})}}
if $pq$ is odd.

$\theta (\tau)$ is related to the Dedekind function
$\eta (\tau)$ by means of the Gauss identity \kacalgebras
\eqn\gaussid{\theta (\tau) = {\eta^2 ({\tau +1\over 2})\over
\eta (\tau +1)},}
which allows one to express the multiplier system $e^{2\pi i\phi(\alpha)}$
in terms of the multiplier system of the Dedekind function. The latter
involves the {\it Dedekind symbol} $S(p;q)$
\ref\slang{See, for example: S. Lang,
``Introduction to Modular Forms'', Springer-Verlag,
New York 1976.} which is defined for relatively prime numbers $p,q$
as follows:
\eqn\dedekind{S(p;q) \equiv \sum_{n=1}^{q-1} ((n/q -1/2)) ((np/q -1/2)),}
with $((x))\equiv x$ modulo integers and $-1/2 \leq ((x)) \leq 1/2$.
One derives in this way
\eqn\expliciteven{e^{2\pi i\theta (p;q)} =
e^{2\pi i(1/2 S(p;q) - S(p+q;2q))}}
if $pq$ is even, and
\eqn\explicitodd{e^{2\pi \theta (p;q)} = e^{2\pi i\theta (p+q;q)}=
e^{2\pi i(1/2 S(p;q) - S(p;2q))}}
if $pq$ is odd. (Notice that if $p,q$ are coprime $p+q,q$ are
coprime too.)

\listrefs
\bye